\documentclass[aps,prd,twocolumn,superscriptaddress,nofootinbib,longbibliography]{revtex4-2}
\pdfoutput=1

\usepackage[T1]{fontenc}
\usepackage[utf8]{inputenc}
\usepackage{graphicx}
\usepackage{amsmath,amssymb,amsfonts}
\usepackage[svgnames]{xcolor}
\usepackage{slashed}
\usepackage{bm}
\usepackage{comment}
\usepackage[colorlinks=true,linkcolor=blue,citecolor=blue,urlcolor=blue]{hyperref}

\allowdisplaybreaks
\usepackage[normalem]{ulem}

\begin{document}

\title{Fire at the Tip of the Throat: Hagedorn Phase after Brane–antibrane Inflation?}

\author{Dibya Chakraborty}
\email{dibyac@fisica.ugto.mx}
\affiliation{School of Physics,
Indian Institute of Science Education and Research \\ Thiruvananthapuram,
Thiruvananthapuram 695551, India}

\author{Ahmed Rakin Kamal}
\email{ahmedrakinkamaltunok@gmail.com}
\affiliation{Department of Theoretical Physics and Astrophysics, Faculty of Science, Masaryk University, Kotl\'a\v{r}sk\'a 2, CS-61137 Brno, Czechia}
\affiliation{Department of Mathematics and Natural Sciences, BRAC University, Kha 224, Bir Uttam Rafiqul Islam Avenue, Dhaka 1212, Bangladesh}

\begin{abstract}
We study the post-inflationary open-string Hagedorn phase in perturbatively moduli stabilized brane-antibrane inflation. In this class of models, the volume modulus is stabilized by perturbative corrections rather than by non-perturbative effects to the superpotential, thereby avoiding the standard brane-antibrane $\eta$-problem. Since inflation ends through tachyon condensation and brane-antibrane annihilation, the endpoint is intrinsically stringy and need not be described immediately by an ordinary radiation bath. We analyze whether the energy released at annihilation can drive the visible sector into an open-string Hagedorn phase, and study the consequences for dark radiation. If the Standard Model (SM) branes lie in the same throat as where the annihilation occurs, we find that a modest fraction of the released energy deposited into surviving visible open strings is sufficient to enter the Hagedorn regime and can suppress the effective number of relativistic species $\Delta N_{\rm eff}$ below current observational bounds. If the SM lies in a different throat, the result depends on inter-throat energy transfer: prompt or delayed transfer can still yield a visible Hagedorn phase. However, the mechanism is most efficient when the local string scale in the SM throat is lower than or comparable to that in the annihilation throat. 
\end{abstract}

\maketitle

\section{Introduction}
Inflationary models which are string inspired often ends in a regime where the low-energy effective field theory (EFT) description becomes insufficient (see \cite{Kallosh:2025ijd, Cicoli:2026bqo} for a recent review on the status of inflationary cosmology). Brane-antibrane inflation \cite{Burgess:2001fx} is a particularly sharp example: unlike ordinary single-field models, where reheating is usually modeled by perturbative inflaton decay, D3/$\overline{\mathrm{D3}}$ inflation ends when the open string stretched between the brane and antibrane becomes tachyonic \cite{Sen:1998sm, Sen:1999mg, Sen:1999xm} and the pair annihilates. The end of inflation is therefore intrinsically stringy. This raises a simple but important question: should the post-inflationary state be described immediately as ordinary radiation, or does it first pass through a high-temperature string phase? 

Recent progress in perturbatively stabilised brane-antibrane inflation \cite{Cicoli:2024bwq} makes this question sharper. The basic idea of brane-antibrane inflation is that the relative position of branes in the compact space can play the role of an inflaton \cite{Burgess:2001fx}. In the warped D3/$\overline{\mathrm{D3}}$ realisation of KKLMMT \cite{Kachru:2003sx}, a mobile D3-brane moves in a warped throat towards an $\overline{\mathrm{D3}}$-brane sitting at the tip, and the separation between brane-antibrane $r$ acts as the inflaton. The four-dimensional potential contains the warped antibrane tension and the attractive Coulomb interaction,
\begin{equation}
    V_{\rm inf}(r,\mathcal V)=C_A\left(1-\frac{D_A}{(rM_{KK})^4}\right),
\end{equation}
where $C_A$ is the warped antibrane-tension contribution, $\mathcal{V}$ is the volume of the 6D Calabi-Yau (CY) manifold, $M_{KK}$ is the lowest non-zero eigenvalue of the Kaluza-Klein mass tower associated with the compactification, and finally $D_A$ is associated with the warping of the throat. In conventional KKLT, the K\"ahler modulus is fixed by non-perturbative effects in the superpotential, generated for instance by Euclidean D3-brane instantons or gaugino condensation on D7-branes \cite{Kachru:2003aw, Balasubramanian:2005zx, McAllister:2024lnt, McAllister:2025qwq}. The resulting non-perturbative superpotential generically depends on the mobile D3-brane position, and this induces an inflaton mass of the order the Hubble scale $(H)$, leading to the familiar $\eta$-problem \cite{Kachru:2003sx}. Although slow-roll inflation can be recovered after additional tuning, explicit constructions have realised it near an approximate inflection point rather than with the original inverse-power Coulombic potential \cite{Baumann:2007np,Baumann:2007ah}.

The construction of Ref.~\cite{Cicoli:2024bwq} avoids this obstruction by stabilizing the volume modulus purely through perturbative corrections\footnote{Inflationary applications of perturbative moduli stabilization have also been studied in \cite{Antoniadis:2018hqy,Hai:2025pki,Chakraborty:2025wqn, Bera:2024ihl, Bera:2024zsk, Chakraborty:2025yms, Basiouris:2021sdf, Basiouris:2025yir}.}. In this framework, the scalar potential for the K\"ahler moduli is generated by known perturbative effects, including the leading $\mathcal{O}(\alpha'^3)$ correction to the K\"ahler potential \cite{Becker:2002nn}, one-loop  $\mathcal{O}(\alpha'^3)$ correction \cite{Antoniadis:2018hqy, Antoniadis:2019rkh} and 1-loop redefinitions of 4-cycle sizes \cite{Conlon:2009kt, Conlon:2010ji}. For a K\"ahler modulus $\tau\,(T=\tau+i \theta)$ and $\theta$ being the axion, this generates a dS minima for certain values of the parameters. The F-term potential in terms of the K\"ahler potential $(K)$ and superpotential $(W)$ is of the form
\begin{equation}
\frac{V}{3 W_0^2}= e^K \left(\frac13 K^{T\overline{T}}K_TK_{\overline{T}} -1 \right)\simeq \frac{\alpha}{\tau^4} - \frac{\xi \sqrt{g_s}}{4 c \tau^{9/2}}\left(\ln\tau  -\frac{c}{g_s^2}\right)\,.
\end{equation}
where $W_0$ is the tree-level flux superpotential and $g_s$ is the string coupling. $\xi$ is related to the Euler characteristics of the CY and $\zeta(s)$ denotes the Riemann zeta function, and
$c\equiv\zeta(3)/\pi^3\simeq0.04$ is a dimensionless coefficient,
with the normalisation adopted as in Ref.~\cite{Cicoli:2024bwq} and $\alpha$ being the moduli redefinition parameter. The string-loop corrections \cite{Berg:2005ja,Berg:2007wt,Cicoli:2007xp}, and higher-derivative \(F^4\) corrections to the four-dimensional effective action \cite{Ciupke:2015msa} have also been checked to be sub-leading and do not ruin the minima of the potential \cite{Cicoli:2024bwq}. The perturbative corrections to the K\"ahler potential stabilizes the overall CY volume $\sigma=\mathcal{V}^{2 / 3}=\tau-\frac{1}{6}\left(M_{\mathrm{KK}} r\right)^2$. The relevant point is that the perturbative stabilization potential is formulated in terms of the physical volume combination $\sigma$, which remains close to its stabilized value as the mobile D3-brane evolves. This ensures that the brane-antibrane scenario does not have an $\eta$-problem which is the case for moduli stabilisation using non-perturbative stabilisation corrections \cite{Kachru:2003sx,Burgess:2022nbx}. \par

Reheating after brane-antibrane annihilation in warped compactifications has been studied extensively. The possibility that sectors localised in distinct warped throats can communicate through the tunnelling of bulk modes was first analysed in \cite{Dimopoulos:2001qd,Dimopoulos:2001ui}. These works established the basic mechanism of inter-throat energy transfer, although they did not consider a Hagedorn phase arising after inflation. Their ideas were subsequently applied to reheating after brane-antibrane inflation. Ref. \cite{Barnaby:2004gg} showed that a significant fraction of the annihilation energy can be deposited in a Standard Model throat even when it is separated from the inflationary throat. \cite{Kofman:2005yz} described the post-annihilation cascade from tachyon condensation through highly excited closed strings and throat-localised KK modes to light Standard Model degrees of freedom. \cite{Chialva:2005zy} compared the relevant decay and tunnelling time scales in a multi-throat geometry , while  \cite{Langfelder:2006vd} emphasised the dependence of the transfer rate on the KK wavefunctions and on the UV gluing of the throats to the bulk compactification. These results motivate the effective description of inter-throat transfer adopted in Sec.~IV in terms of the rates $\Gamma_{A\to S}$ and $\Gamma_{\rm tot}$.

A different side of the story comes from the study of Hagedorn phase in string cosmology\footnote{See \cite{McAllister:2007bg,Baumann:2014nda,Kallosh:2018zsi,Cicoli:2023opf} for a recent review on string cosmology and string inflation.}. The exponential growth of the perturbative string density of states,
\begin{equation} 
\Omega(E)\sim e^{\beta_H E}, 
\end{equation} defines the Hagedorn temperature $T_H=\beta_H^{-1}$. Near this temperature, additional energy is stored predominantly in oscillator excitations and in the length of highly excited strings rather than in an increasing temperature. The thermodynamics of fundamental and open strings in this regime, including its microcanonical description and finite-volume behaviour, has been studied in Refs.~\cite{Atick:1988si,Deo:1989bv,Lee:1997iz,Abel:1999rq}. This literature should be distinguished from string-gas cosmology, where a thermal string gas is assumed to participate directly in the background cosmological evolution and in the dynamics of the compact dimensions \cite{Brandenberger:1988aj,Tseytlin:1991xk}.

In string compactifications, a closely related cosmological problem is the generic presence of light hidden-sector degrees of freedom: such as axions, hidden gauge sectors, and other weakly coupled relativistic species. If these are populated during reheating, they contribute to the effective number of neutrino species $\Delta N_{\rm eff}$, and can easily conflict with observational bounds. Ref.~\cite{Frey:2005jk,Frey:2021jyo} pointed out that an open-string Hagedorn phase in the visible sector provides a natural way to suppress this contribution: if the Standard Model degrees of freedom live on branes and enter a Hagedorn phase, while the dark-radiation sector does not participate in the same thermal plasma; the enormous visible sector entropy enhances the effective visible number of degrees of freedom at reheating and reduces the final ratio between energy densities of dark radiation and visible sector, $\rho_{\rm dr}/\rho_{\rm vis}$. In this way, the Hagedorn phase is not merely a formal feature of finite-temperature string theory, but can have direct cosmological consequences for dark radiation in warped compactifications.

Moreover, there has been a lot of recent progress in understanding the Hagedorn phase and its subsequent gravitational wave production \cite{Frey:2021jyo, Frey:2024jqy, Frey:2023Boltzmann}. In addition, progress has been made in understanding cosmic superstrings and also their implications for gravitational waves \cite{Brunelli:2025ems, Brunelli:2025eif, Brunelli:2026qkp, Conlon:2024uob, Conlon:2025mqt, SanchezGonzalez:2025uco,Villa:2026pnb, Ghoshal:2025tlk, Revello:2024rxn, Oikonomou:2023bli, Oikonomou:2023qfz, Datta:2024bqp, Blanco-Pillado:2024aca}. Taken together, these developments suggest the need to study controlled brane-antibrane inflation which provides a natural setting in which both a Hagedorn phase and cosmic-superstring \cite{Burgess:2001fx, Burgess:2001vr, Barnaby:2004gg, Majumdar:2003da, Majumdar:2002hy, Sen:1998tt} relics may arise, with potentially observable consequences for dark radiation and gravitational waves. \par

Our main result is that a post-inflationary open-string Hagedorn phase can arise naturally after brane-antibrane annihilation, provided that a sufficient fraction of the released energy is deposited into surviving visible open-string degrees of freedom. When the Standard Model sector is localised in the same warped throat as the annihilating brane-antibrane pair, the condition for entering the Hagedorn regime is parametrically mild: for weak string coupling, only a few to ten percent of the annihilation energy needs to be transferred to a modest visible brane stack. The resulting Hagedorn epoch is typically short, lasting ${\cal O}(1)$--few e-folds, but the associated large visible-sector entropy can substantially suppress the contribution of decoupled relativistic species to $\Delta N_{\rm eff}$.

When the Standard Model instead resides in a different warped throat, the outcome depends both on the relative amount of warping in the two throats and on how efficiently the annihilation energy is transferred between them. Denoting by $M_{s,A}$ and $M_{s,S}$ the local warped string scales in the annihilation and Standard Model throats, respectively, we find that Hagedorn reheating is favoured when the Standard Model throat is at least as strongly warped as the annihilation throat, corresponding roughly to $M_{s,S}\lesssim M_{s,A}$. If the inter-throat transfer is prompt, the annihilation energy is deposited before significant cosmological dilution. If instead the transfer is delayed, the energy density decreases before reaching the visible sector, and the transfer must therefore occur sufficiently early to remain above the Hagedorn threshold. Writing $\Gamma_{\rm tot}$ for the total depletion rate of the annihilation-throat energy reservoir, this requires it to exceed a minimum Hagedorn-entry rate $\Gamma_{\min}$ while remaining below the Hubble scale at the end of inflation, $H_{\rm end}$. For the benchmark values considered in this work, such a delayed Hagedorn window exists for $M_{s,S}/M_{s,A}\lesssim 1.7$ and becomes progressively broader as the Standard Model throat is more strongly warped. Thus, in both the prompt and delayed regimes, the visible Hagedorn phase is most easily realised when the local string scale of the Standard Model throat is lower than, or at most comparable to, that of the annihilation throat.

The paper is organised as follows. In Sec.~II we review the open-string Hagedorn criterion and its implications for $\Delta N_{\rm eff}$. In Sec.~III we study the case in which the Standard Model sector and the annihilating brane-antibrane pair are localised in the same throat. In Sec.~IV we analyse the different-throat scenario, considering both prompt and delayed inter-throat transfer. We conclude in Sec.~V.

\section{Criteria for Hagedorn Phase and $\Delta N_{eff}$ supression}

In warped compactifications \cite{Giddings:2001yu} with a Klebanov-Strassler throat \cite{Klebanov:1998hh, Klebanov:2000hb,Klebanov:2000nc}, the Hagedorn temperature represents the local string scale in a warped throat X \footnote{In our notation, X=S for the SM throat and X=A for the brane-antibrane inflation throat.}
\begin{equation}
    T_{H,X} =q_H M_{s,X}
\end{equation}
where $q_H$ is an order-one constant that depends on the convention used to define the string mass $M_s$ \footnote{In our convention, the string length is given by $\ell_s=2\pi\sqrt{\alpha'}$, $M_s=\ell_s^{-1}$ and one has $\displaystyle q_H=\frac{1}{\sqrt{2}}$ \cite{ValeixoBento:2025emh}.}. In terms of energy density, the condition for a Hagedorn phase is roughly
\begin{equation}
    \rho_{\text{local}}>M_{s,X}^4, 
\end{equation}
where the subscript local stands for the local energy density in the warped throat. If $M_{s,X}\ll M_p$ where $M_p$ is the Planck mass, then using Friedmann equations, the condition on the Hubble parameter ($H=\dot{a}/a)$, where $a$ is the scale factor) associated with the energy scale of inflation becomes
\begin{equation}
    H\ll M_{s,X},
\end{equation}
which says that the EFT can be under control. Ref.~\cite{Frey:2021jyo} parametrises the dark-radiation constraint in terms of the ratio between the dark and visible energy densities at neutrino decoupling $t_{\nu}$ as:
\begin{equation}
\Delta N_{\mathrm{eff}}=\frac{43}{7}\frac{\rho_{\mathrm{dr}}(t_\nu)}
{\rho_{\mathrm{vis}}(t_\nu)} ,
\end{equation}
where $\rho_{\mathrm{dr}}$ is the energy density in decoupled relativistic species, and $\rho_{\mathrm{vis}}$ denotes the visible-sector radiation energy density. String compactifications generically contain light hidden-sector degrees of freedom, such as axions, hidden gauge sectors, and other weakly coupled fields, which may contribute to $\rho_{\mathrm{dr}}$ and hence to $\Delta N_{\mathrm{eff}}$. If reheating deposits fractions of the available energy in visible radiation $B_{\mathrm{vis}}$ and dark radiation $B_{\mathrm{dr}}$ respectively, then at reheating
\begin{equation} 
\frac{\rho_{\mathrm{dr}}^{\mathrm{rh}}}{\rho_{\mathrm{vis}}^{\mathrm{rh}}}=\frac{B_{\mathrm{dr}}}{B_{\mathrm{vis}}}.
\end{equation}
After reheating, the subsequent evolution of this ratio is controlled by entropy conservation in the visible sector. For an ordinary relativistic plasma, one obtains
\begin{equation}\label{Neff1}
\Delta N_{\mathrm{eff}}=\frac{43}{7}\frac{B_{\mathrm{dr}}}{B_{\mathrm{vis}}}\left(\frac{g_{\star s}(T_\nu)}{g_{\star s}(T_{\mathrm{rh}})}\right)^{1/3},
\end{equation}
where $g_{\star s}(T)$ denotes the effective number of visible-sector entropy degrees of freedom at temperature $T$ and $T_{\rm rh}$ denotes the reheating temperature. Hence, for a fixed branching ratio $B_{\mathrm{dr}}/B_{\mathrm{vis}}$, a large visible-sector entropy at reheating suppresses the final contribution to $\Delta N_{\mathrm{eff}}$.

Ref.~\cite{Frey:2021jyo} proposed that this suppression can be substantially enhanced if the visible sector is reheated to an open-string Hagedorn phase. We use the D3-brane setup of Ref.~\cite{Frey:2021jyo}, in which the visible open strings live on the three large spatial dimensions of a stack of $N$ number of $D3$ branes. In this case the onset of the Hagedorn regime may be expressed as a local four-dimensional energy-density condition,
\begin{equation}\label{energy density condition}
    \rho_{\rm vis}\gtrsim \kappa_H N^2 M_{s,S}^4 ,
\end{equation}
where $M_{s,S}$ is the local warped string scale of the SM throat and $\kappa_H$ is an order-one coefficient which encodes the precise convention for the onset of the Hagedorn regime.\footnote{If the visible sector is realised on D7-branes, Eq.~ \eqref{micro_entropy} must be modified. The open-string gas then propagates not only along the three large spatial dimensions, but also along the internal four-cycle wrapped by the D7-branes. If this four-cycle has local volume $V_4^{\rm loc}$, it is useful to define $\mathcal V_4^{\rm loc}\equiv V_4^{\rm loc}/\ell_{s,S}^4$. The D3 threshold in Eq.~\eqref{energy density condition} is then replaced schematically by $\rho_{\rm vis}\gtrsim \kappa_H N^2\mathcal V_4^{\rm loc}M_{s,S}^4$, up to order-one factors. More generally, the Hagedorn thermodynamics depends on the D-brane dimensionality and on the finite-volume regime of the compact directions \cite{Abel:1999rq,Frey:2021jyo,Frey:2023Boltzmann}. The D7 case reduces effectively to the D3-like estimate when the wrapped four-cycle is effectively close to string-scale, $\mathcal V_4^{\rm loc}\sim 1$.} This threshold follows from the D3 open-string Hagedorn thermodynamics. The microcanonical entropy $S_0$ associated with the system is \cite{Lee:1997iz, Abel:1999rq}
\begin{equation}\label{micro_entropy}
S_o(E)=\beta_H E+\sqrt{\frac{8N^2V_\parallel E}{m\mu^2V_\perp}},
\end{equation}
where $V_\parallel$ is the spatial volume along the D3-branes, $V_\perp$ is the effective transverse volume explored by the long strings, $\mu=(2\pi\alpha')^{-1}$ is the string tension, and $m$ is an order-one parameter. Here \(N\) should be interpreted as the effective Chan--Paton multiplicity of the visible open-string sector participating in the Hagedorn gas. The first term is the Hagedorn entropy, while the square-root correction controls the approach to $T_H=\beta_H^{-1}$. 

It is useful to express the visible-sector energy density in the standard cosmological form as:
\begin{equation}
\rho_{\rm vis}=\frac{\pi^2}{30}g_{\star E}(T)T^4 ,
\end{equation}
where $g_{\star E}(T)$ is defined as the effective number of visible-sector energy degrees of freedom at temperature $T$, in analogy to the usual relativistic-plasma relation $\rho_r=(\pi^2/30)g_\star(T)T^4$ \cite{Baumann:2022mni}. Introducing
\begin{equation}
\epsilon_H\equiv 1-\frac{T_{\rm rh}}{T_H},
\end{equation}
with $\epsilon_H\ll1$ denoting reheating close to the Hagedorn temperature\footnote{Please note that the quantity $\epsilon_H$ should not be confused with Hubble slow-roll parameter.}, Ref.~\cite{Frey:2021jyo} finds 
\begin{equation}
g_{\star E}^{\rm Hag}(T_{\rm rh})\simeq\frac{30N^2}{\pi^4mV_{\perp}}\frac{1}{\epsilon_H^2},
\end{equation}
where $V_{\perp}$ parametrises the effective transverse-volume factor. Equivalently, near $T_H\simeq q_HM_{s,S}$,
\begin{equation}\label{eq:hagedorn_energy_density}
\rho_{\rm vis}^{\rm Hag}
\simeq
\frac{N^2q_H^4}{\pi^2mV_{\perp}}
\frac{M_{s,S}^4}{\epsilon_H^2}.
\end{equation}
So, Eq.~\eqref{energy density condition} corresponds to the onset of the Hagedorn regime, with the order-one factors absorbed into $\kappa_H$. Since near $T_H$, the entropy density is dominated by the Hagedorn term, $s_o\simeq \beta_H\rho_{\rm vis}$, where $s_0$ is the entropy density, one has the following
\begin{equation}
g_{\star s}^{\rm Hag}(T_{\rm rh})\simeq\frac{3}{4}g_{\star E}^{\rm Hag}(T_{\rm rh}).
\end{equation}
Thus, \eqref{Neff1} becomes
\begin{equation}
\Delta N_{\rm eff}\simeq\frac{43}{7}\frac{B_{\rm dr}}{B_{\rm vis}}\left(\frac{4}{3}\right)^{4/3}\left(\frac{g_{\star E}(T_\nu)}{g_{\star E}^{\rm Hag}(T_{\rm rh})}\right)^{1/3}.
\end{equation}
Using the standard value $g_{\star E}(T_\nu)\simeq10.75$ at neutrino decoupling \cite{Dodelson:2003ft, Baumann:2022mni, Husdal:2016haj}, this becomes
\begin{equation}\label{Delta Neff same throat}
\Delta N_{\rm eff} \simeq 29.5\,\frac{B_{\rm dr}}{B_{\rm vis}}\,(mV_{\perp})^{1/3}N^{-2/3}\epsilon_H^{2/3}.
\end{equation}
Thus, in the D3-brane Hagedorn regime, reheating closer to $T_H$ yields a very small $\epsilon_H$ and therefore a smaller $\Delta N_{\rm eff}$. The duration of the Hagedorn epoch can be estimated from the equation of state of the long-string gas. In the Hagedorn regime the temperature remains approximately pinned near \(T_H\), and additional energy is stored mainly in the length and oscillator excitations of highly excited strings. In string-gas cosmology this phase is treated as having negligible pressure $p_H\simeq0$ \cite{Brandenberger:1988aj,Tseytlin:1991xk,Deo:1989bv,Nayeri:2005ck, Brandenberger:2011et, Chen:2007js}. Thus, the Hagedorn energy density $(\rho_H)$ redshifts approximately as matter,
\begin{equation}
    \dot\rho_H+3H\rho_H\simeq0,\qquad\rho_H\propto a^{-3}.
\end{equation}
For the D3-brane open-string Hagedorn gas discussed above, the energy density scales as
\begin{equation}
    \rho_H \propto \frac{1}{\epsilon_H^2},\qquad\epsilon_H\equiv 1-\frac{T}{T_H}.
\end{equation}
If the Hagedorn phase begins with $\epsilon_H=\epsilon_{H,i}\ll1$ and ends when the system exits the stringy regime, $\epsilon_H=\epsilon_{H,\rm exit}\simeq 1$ i.e. $T\ll T_H$, then
\begin{equation}
    \frac{\rho_i}{\rho_{\rm exit}}\simeq\left(\frac{\epsilon_{H,\rm exit}}{\epsilon_{H,i}}\right)^2 .
\end{equation}
Using $\rho_H\propto a^{-3}$, the number of e-folds spent in the Hagedorn phase is therefore
\begin{equation}
    \Delta N_H\simeq\frac{1}{3}\ln\left(\frac{\rho_i}{\rho_{\rm exit}}\right)\simeq\frac{2}{3}\ln\left(\frac{\epsilon_{H,\rm exit}}{\epsilon_{H,i}}\right).
\end{equation}
We thus take $\epsilon_{H,\rm exit}\sim1$, giving
\begin{equation}\label{Duration}
    \Delta N_H\simeq\frac{2}{3}\ln\left(\frac{1}{\epsilon_{H,i}}\right).
\end{equation}
Thus, a reheating event very close to the Hagedorn temperature, $\epsilon_{H,i}\ll1$, leads to a longer intermediate stringy phase before the system enters an ordinary radiation bath. With this in mind, we now explore two possible scenarios in detail:
\begin{itemize}
    \item When the SM and the Inflation are in the same throat i.e $X=S=A$. 
    \item When the SM and the Inflation are in different throats i.e $S\neq A$. 
\end{itemize}

\section{SM in the same throat}
In the same-throat scenario we identify the annihilation throat with the visible throat \footnote{
By ``SM in the same throat'' we do not mean that the Standard Model lives on the annihilating D3/\(\overline{\mathrm{D3}}\) pair itself. That would be problematic, since this pair disappears through tachyon condensation. Rather, we mean that the visible sector is supported on a surviving spectator sector in the same warped region: for example, fractional D3-branes at a local orbifold/del Pezzo singularity, D7-branes wrapping a four-cycle which enters the throat, or an additional brane stack separated from the annihilating pair but still local to the same IR region. Such local D-brane realisations of chiral gauge sectors and SM-like quivers are standard in bottom-up string model building \cite{Aldazabal:2000sa,Krippendorf:2010hj,Cicoli:2021dhg}, while D7-brane embeddings in conifold-like throats are also well studied \cite{Ouyang:2003df}. The assumption is that this spectator visible sector is not tachyonically connected to the annihilating D3/\(\overline{\mathrm{D3}}\) pair, survives the annihilation, and is close enough in the local throat geometry to receive a fraction of the released stringy energy. Under these conditions the same-throat case is consistent: it describes local visible reheating into surviving open-string degrees of freedom, not the reheating of the unstable brane-antibrane pair itself.
}
\begin{equation*}
   A=S. 
\end{equation*}
The brane-antibrane potential in the warped throat can be written as \cite{Cicoli:2024bwq}
\begin{equation}
\begin{gathered}
V_{\mathrm{inf}}(r,\mathcal V)=C_A\left(1-\frac{D_A}{(rM_{KK})^4}\right),\\
C_A=\frac{M_p^4}{4\pi\mathcal V^{2}}e^{\varrho_A},\\
D_A=\left(\frac{3}{4\pi}\right)^3\mathcal{V}^{-2/3} e^{\varrho_A}.
\end{gathered}
\label{eq:warped_ba_potential}
\end{equation}
Here $e^{\varrho_A}$ is the warp factor associated with the annihilation throat i.e. the strongly warped throat and the warp factor $h$ over the compact manifold is defined as \cite{Giddings:2005ff,ValeixoBento:2025emh,Cicoli:2024bwq,Agarwal:2025rqd}
\begin{equation}
h=1+\frac{e^{-4A}}{\mathcal{V}^{2/3}},
\end{equation}
where we work in the strongly warped limit $e^{-4A}\gg \mathcal{V}^{2/3}$ where the second term dominates and becomes:
\begin{equation}
    e^{-\varrho_A}\approx \frac{e^{-4A}}{\mathcal{V}^{2/3}}\simeq \frac{1}{\mathcal{V}^{2/3}e^{-\frac{8\pi K}{3 g_s M}}}
\end{equation}
The quantity $C_A$, as previously introduced, is the warped brane-antibrane tension contribution to the four-dimensional potential. The energy density available at the onset of brane-antibrane annihilation is well approximated by the constant term,\footnote{
More precisely, the energy density available at the onset of annihilation is
\begin{equation*}
    \rho_{\rm ann}^{(A)}=\chi_A C_A, \qquad \chi_A=1-\Delta_C(r_{\rm tach}),
\end{equation*}
where \(r_{\rm tach}\) denotes the separation at which the stretched open string becomes tachyonic. Thus \(\rho_{\rm ann}^{(A)}\simeq C_A\) should not be understood as an extrapolation of the Coulomb term to \(r\to0\). The Coulomb term is a long-distance supergravity approximation and ceases to be valid once the brane-antibrane separation becomes string scale and tachyon condensation takes over \cite{Burgess:2001fx}. In the warped throat, writing
\begin{equation*}
    V_A(r)=C_A[1-\Delta_C(r)],
\quad
\Delta_C(r)=
\frac{27\,e^{\varrho_A}}{32\pi^2T_3r^4},\quad M_s=\frac{g_s^{1/4}}{\sqrt{4\pi \mathcal{V}}}M_p
\end{equation*}
the largest value suggested by the leading Coulomb expression is obtained at the tip \(r_0\). Using \cite{Cicoli:2024bwq},
\begin{equation}
    e^{-\varrho_A}=\left(\frac{R}{r_0}\right)^4,
\qquad R^4=\frac{27}{8(2\pi)^3}\frac{g_sM_AK_A}{M_s^4}, \qquad T_3=\frac{2\pi}{g_s}M_s^4,
\end{equation}
where $R$ is the characteristic length of the CY. The previous identity leads to
\[
\Delta_C(r_0)=\frac{27}{32\pi^2 T_3R^4}
=
\frac{1}{M_AK_A}.
\]
Hence, since \(r_{\rm tach}\gtrsim r_0\),
\[
\chi_A
=
1-\Delta_C(r_{\rm tach})
\gtrsim
1-\frac{1}{M_AK_A}.
\]
For the flux choices of Ref.~\cite{Cicoli:2024bwq}, for example \(M_AK_A=230\)--\(700\), this gives \(\chi_A\simeq0.996\)--\(0.999\). Thus \(\rho_{\rm ann}^{(A)}\simeq C_A\) is an appropriate enough leading approximation in the warped regime.
}
\begin{equation}
\rho_{\rm ann}^{(A)}
\simeq
C_A.
\label{eq:rho_ann_same_throat}
\end{equation}
Local warped string scale in the annihilation throat is \cite{Cicoli:2024bwq}
\begin{equation}
M_{s,A}
=
\frac{g_s^{1/4}}{\sqrt{4\pi \mathcal V}}
e^{\frac{\varrho_A}{4}}M_p,
\end{equation}
Combining this with the expression for \(C_A\) gives
\begin{equation}
C_A=\frac{4\pi}{g_s}M_{s,A}^4.
\label{eq:CA_Ms_relation}
\end{equation}
Therefore the annihilation energy density is above the local string scale \footnote{The condition to satisfy is that $g_s <4\pi$, which is automatically satisfied in perturbative string theory.},
\begin{equation}
\rho_{\rm ann}^{(A)}\simeq\frac{4\pi}{g_s}M_{s,A}^4.
\label{eq:rho_ann_Ms}
\end{equation}

We now introduce the fractions of the annihilation energy that are deposited into the visible open-string sector and into dark radiation:
\begin{equation}
B_{\rm vis}=\frac{\rho_{\rm vis,open}^{(A)}}{C_A},\qquad B_{\rm dr}=\frac{\rho_{\rm dr}^{(A)}}{C_A}.
\label{eq:zeta_definitions}
\end{equation}
Here \(\rho_{\rm vis,open}^{(A)}\) denotes the energy density deposited into surviving visible-sector open strings in the annihilation throat, while \(\rho_{\rm dr}^{(A)}\) denotes the energy density deposited into dark radiation. This implies
\begin{equation}
B_{\rm vis}+B_{\rm dr}+B_{\rm other}=1,
\end{equation}
where $B_{\rm other}$ accounts for energy stored in other channels, such as closed strings, KK modes, or cosmic-string production \cite{Burgess:2001vr, Majumdar:2002hy, Firouzjahi:2005dh, Davis:2008kg, Copeland:2009ga}.
The branching ratio entering the $\Delta N_{\rm eff}$ is then
\begin{equation}
\frac{B_{\rm dr}}{B_{\rm vis}}=\frac{\rho_{\rm dr}^{\rm rh}}
{\rho_{\rm vis}^{\rm rh}}
\label{eq:branching_same_throat}
\end{equation}
The visible open-string energy density is
\begin{equation}
\rho_{\rm vis,open}^{(A)}=B_{\rm vis} C_A=B_{\rm vis}\frac{4\pi}{g_s}M_{s,A}^4.
\label{eq:visible_energy_same_throat}
\end{equation}

The condition for the visible sector to enter an open-string Hagedorn phase is that this energy density exceed the local string-density threshold \footnote{$\kappa_H \propto \frac{1}{\epsilon_{H,exit}^2}\sim 1$ so for Hagedorn phase, density should be more than $\kappa_H N^2 M_{s,A}^4$.},
\begin{equation}
\rho_{\rm vis,open}^{(A)}\gtrsim \kappa_H N^2 M_{s,A}^4,
\label{eq:hagedorn_threshold_same_throat}
\end{equation}
where \(N^2\) is the Chan--Paton multiplicity of the visible brane stack and \(\kappa_H=O(1)\) parametrises the precise threshold convention.  Substituting Eq.~\eqref{eq:visible_energy_same_throat} into Eq.~\eqref{eq:hagedorn_threshold_same_throat}, we obtain the condition for the visible sector in the annihilation throat to enter the open-string Hagedorn regime 
\begin{equation}
B_{\rm vis}\gtrsim\frac{\kappa_H N^2 g_s}{4\pi}.
\label{eq:same_throat_hagedorn_condition}
\end{equation}
For weak string coupling this is a mild requirement. Taking \(g_s=0.05\) and \(\kappa_H=1\), one finds
\begin{equation}
\begin{array}{c|c}
N & \text{minimum visible fraction} \\ \hline
1 & B_{\rm vis}\gtrsim 0.4\% \\
2 & B_{\rm vis}\gtrsim 1.6\% \\
3 & B_{\rm vis}\gtrsim 3.6\% \\
5 & B_{\rm vis}\gtrsim 10.0\% \\
7 & B_{\rm vis}\gtrsim 19.5\%
\end{array}
\end{equation}
Hence, even for a moderately large visible brane sector, the same-throat Hagedorn condition requires only an order \(10\%\) fraction of the annihilation energy to enter surviving visible open strings.
In other words, for a modest visible brane stack, only a few to ten percent of the annihilation energy must be deposited into surviving visible open strings for the same-throat sector to enter the Hagedorn regime. 

Using the same-throat Hagedorn condition,
\begin{equation}
    B_{\rm vis}\gtrsim B_{\rm vis}^{\rm min}\equiv\frac{\kappa_HN^2g_s}{4\pi},
\end{equation}
we can translate the dark-radiation constraint into a bound on \(B_{\rm dr}\). We start by rewriting eq.~\eqref{eq:hagedorn_energy_density} as
\begin{equation}
    \epsilon_H^2=\frac{N^2q_H^4g_s}{4\pi^3mV_{\perp}\,B_{\rm vis}},
\end{equation}
which modifies the Hagedorn expression for \(\Delta N_{\rm eff}\) from Eq.~\eqref{Delta Neff same throat} to
\begin{equation}
    \Delta N_{\rm eff}\simeq 29.5 \frac{B_{\rm dr}}{B_{\rm vis}}\left(\frac{q_H^4g_s}{4\pi^3B_{\rm vis}}\right)^{1/3}.
\end{equation}
For \(q_H=1/\sqrt2\), \(g_s=0.05\) and imposing \(\Delta N_{\rm eff}<0.3\),
\begin{equation}
    B_{\rm dr}\lesssim0.219\,B_{\rm vis}^{4/3}.
\end{equation}
At the minimal visible fraction required for Hagedorn entry, this gives
\begin{equation}
\begin{array}{c|c|c}
N & B_{\rm vis}^{\rm min} & B_{\rm dr}^{\rm max} \\ \hline
1 & 0.4\% & 1.4\times10^{-4} \\
3 & 3.6\% & 2.6\times10^{-3} \\
5 & 10\% & 1.0\times10^{-2} \\
7 & 19.5\% & 2.5\times10^{-2}
\end{array}
\end{equation}
Thus, if the visible sector is only just above the Hagedorn threshold, the allowed dark-radiation branching is at most $1\%$ level. Larger \(B_{\rm vis}\) relaxes the bound as \(B_{\rm dr}^{\rm max}\propto B_{\rm vis}^{4/3}\). So, the annihilation energy can go into dark sectors but not much of it can remain as decoupled relativistic dark radiation. The caveat is important: $B_{\mathrm{dr}}$ here means energy that remains in decoupled relativistic species until late times. Energy initially going into closed strings, KK modes, cosmic strings, or hidden sectors, is not necessarily counted as $\Delta N_{\text {eff }}$ if it becomes massive, decays into visible radiation or annihilates. 

One can similarly estimate how close the visible bath is to the Hagedorn temperature. Equating
\begin{equation}
\rho_{\rm vis,open}^{(A)}=B_{\rm vis}\frac{4\pi}{g_s}M_{s,A}^4
\end{equation}
with the D3 open-string Hagedorn energy density,
\begin{equation}
\rho_{\rm vis}^{\rm Hag}\simeq\frac{N^2q_H^4}{\pi^2mV_{\perp}}\frac{M_{s,A}^4}{\epsilon_H^2},
\end{equation}
gives
\begin{equation}
\epsilon_H=\left(\frac{N^2q_H^4g_s}{4\pi^3mV_{\perp}\,B_{\rm vis}}\right)^{1/2}.
\end{equation}
For \(g_s=0.05\), \(q_H=1/\sqrt2\), \(mV_{\perp}=1\), and taking \(B_{\rm vis}\) at the minimum value required by the Hagedorn condition, one finds
\begin{equation}
\begin{array}{c|c|c}
N & B_{\rm vis}^{\rm min} & \epsilon_H \\ \hline
1 & 0.4\% & 0.16 \\
2 & 1.6\% & 0.16 \\
3 & 3.6\% & 0.16 \\
5 & 10.0\% & 0.16 \\
7 & 19.5\% & 0.16
\end{array}
\end{equation}
The equality of the last column is not accidental at the threshold
\(B_{\rm vis}^{\rm min}=N^2g_s/(4\pi)\), the dependence on \(N\) cancels, and
\begin{equation}
\epsilon_H^{\rm min}=\frac{q_H^2}{\pi\sqrt{mV_{\perp}}}\simeq0.16 .
\end{equation}
For larger visible energy deposition, \(\epsilon_H\) decreases as
\(\epsilon_H\propto B_{\rm vis}^{-1/2}\), so the bath is even closer to \(T_H\). For example, for \(N=3\), \(g_s=0.05\), and \(\kappa_H=1\), one has
\(B_{\rm vis}^{\rm min}\simeq3.6\%\). 
\begin{equation}
\begin{array}{c|c}
B_{\rm vis} & \epsilon_H \\ \hline
3.6\% & 0.16 \\
10\% & 0.095 \\
30\% & 0.055 \\
100\% & 0.030
\end{array}
\end{equation}
Let us now analyze the duration of the Hagedorn phase using Eq.~\eqref{Duration}. For $N=3, g_s=0.05, q_H=1 / \sqrt{2}, m V_{\perp}=1$, $B_{\text {vis }}>0.03$ so
$$
\begin{array}{cccc}
B_{\text{vis }}=1: & \epsilon_H \simeq 0.030 & \Rightarrow & \Delta N_H \simeq 2.3 . \\
B_{\text{vis}}=0.3: & \epsilon_H \simeq 0.055 & \Rightarrow & \Delta N_H \simeq 1.9 . \\
B_{\text{vis}}=0.1: & \epsilon_H \simeq 0.095 & \Rightarrow & \Delta N_H \simeq 1.6 .
\end{array}
$$
The Hagedorn phase is likely a short but real post-inflationary stringy phase: roughly one to a few efolds, depending on visible energy deposition.

\section{SM In Different Throat}
In this scenario, we set the SM to be in a different throat, i.e $A\neq S$ in our notation. In a two-throat realisation, the energy released by D3/\(\overline{\mathrm{D3}}\) annihilation is initially localised in the annihilation throat $A$, while the Standard Model degrees of freedom live in a different throat $S$. The relevant reheating chain is \(D3/\overline{D3}\) annihilation in \(A\rightarrow\) massive closed strings/\(A\)-throat KK modes \(\rightarrow\) inter-throat transfer \(\rightarrow\) visible open strings in \(S\).
Refs.~\cite{Barnaby:2004gg,Chialva:2005zy, Frey:2005jk} argued that reheating in a two-throat brane-inflation setup can proceed through massive closed strings and throat-localised KK modes: the annihilation energy is first deposited in the inflationary throat, after which KK modes can tunnel to another throat and reheat the Standard Model sector before decaying gravitationally, in a suitable range of parameters. Ref.~\cite{Langfelder:2006vd} emphasized that this inter-throat transfer is geometry-dependent. Thus, wavepacket tunneling, overlaps between throat-localised modes, and the details of the UV gluing can substantially affect the rate. We therefore do not assume perfect or instantaneous transfer, but parameterise the process by an effective rate \(\Gamma_{A\to S}\).

In addition, when transferred to the SM throat, the \(A\)-throat energy can be lost into channels that do not reheat the visible sector, such as bulk gravitons, hidden throats, axions, or other decoupled modes. We therefore write
\begin{equation*}
   \Gamma_{\rm tot}=\Gamma_{A\to S}+\Gamma_{\rm loss}, 
\end{equation*}
where \(\Gamma_{\rm tot}\) is the total depletion rate of the \(A\)-throat energy reservoir. The time at which this reservoir is depleted is controlled by \(\Gamma_{\rm tot}\). The fraction of the depleted energy that is transferred to the SM throat is instead controlled by the branching ratio 
\begin{equation*}
    {\rm Br}_{A\to S}=\frac{\Gamma_{A\to S}}{\Gamma_{\rm tot}}.
\end{equation*}
Once the energy has reached throat \(S\), only a further fraction \(B_{\rm vis}^{(S)}\)  must be converted into visible open strings. Thus, the total visible branching fraction is
\begin{equation*}
    B_{\rm vis}={\rm Br}_{A\to S}\,B_{\rm vis}^{(S)}=\left(\frac{\Gamma_{A\to S}}{\Gamma_{\rm tot}}\right)B_{\rm vis}^{(S)}.
\end{equation*}
Equivalently,
\begin{equation*}
    \rho_{\rm vis,open}^{(S)}=B_{\rm vis}\rho_{\rm available}.
\end{equation*}
Thus, \(\Gamma_{\rm tot}\) determines the time of depletion, while \(B_{\rm vis}\) determines the fraction of the available energy that ultimately appears as a visible open-string energy density. \footnote{Microscopic estimates of inter-throat transfer are strongly geometry-dependent. In the Randall--Sundrum-like treatment of Ref.~\cite{Chialva:2005zy}, the KK tunnelling rate scales parametrically as
$\Gamma_{\rm tunn}\sim (mR_A)^4/|z_A|$.
Depending on the throat geometry, tunnelling can occur before competing losses into bulk gravitons, allowing a sizeable branching fraction into the SM throat. The rate is sensitive to the throat lengths, warp factors, KK spectrum and UV gluing \cite{Barnaby:2004gg,Chialva:2005zy,Langfelder:2006vd}; we therefore keep $\Gamma_{A\to S}$ and $\Gamma_{\rm tot}$ as effective parameters.}

This distinction is important because Hagedorn entry is an energy-density condition in the SM throat. The two relevant regimes are therefore:
\begin{equation}
    \Gamma_{\rm tot}= H_{\rm end}\qquad\text{prompt transfer},
\end{equation}
and
\begin{equation}
    \Gamma_{\rm tot}< H_{\rm end}\qquad\text{delayed transfer},
\end{equation}
where $H_{\rm end}$ denotes the Hubble parameter at the end of inflation. In the prompt regime, the \(A\)-throat energy is depleted before significant dilution of energy due to cosmic expansion, so the energy available for deposition into the SM throat is still of order the annihilation energy density, up to branching fractions. In the delayed regime, on the contrary, the energy remains stored in the \(A\)-throat sector until the Hubble scale drops to
\begin{equation}
    H\simeq \Gamma_{\rm tot}.
\end{equation}
Assuming that this sector dominates the energy density until it decays or transfers, the total energy density at deposition is
\begin{equation}
    \rho_{\rm dep}\simeq 3M_p^2\Gamma_{\rm tot}^2<3M_p^2H_{\rm end}^2 .
\end{equation}
Only the fraction \(B_{\rm vis}\rho_{\rm dep}\) is converted into visible open strings. Thus, in the delayed case, the visible Hagedorn condition is controlled not by the original annihilation energy density, but by the diluted density at \(H\simeq\Gamma_{\rm tot}\). Thus, one needs to check if this diluted density is above the required Hagedorn density. 

\subsection{Regime 1: Prompt Transfer}
So, we first deal with the case where the annihilation throat energy is depleted before significant cosmological dilution. In this case, we have
\begin{equation}
\rho_{\mathrm{vis}, \mathrm{open}}^{(S)}=B_{\mathrm{vis}} \rho_{\mathrm{ann}}^{(A)}=B_{\mathrm{vis}} \frac{4 \pi}{g_s} M_{s, A}^4 .
\end{equation}
alongside the Hagedorn condition
\begin{equation}
\rho_{\mathrm{vis}, \mathrm{open}}^{(S)} \gtrsim \kappa_H N^2 M_{s, S}^4 
\end{equation}
gives us
\begin{equation}
B_{\mathrm{vis}} \gtrsim \frac{\kappa_H N^2 g_s}{4 \pi}\left(\frac{M_{s, S}}{M_{s, A}}\right)^4 .
\end{equation}
For \(g_s=0.05\), \(N=3\), and \(\kappa_H=1\), the prompt-transfer Hagedorn condition becomes
\begin{equation}\label{prompt_transfer_table}
\begin{array}{c|c}
M_{s,S}/M_{s,A} & B_{\rm vis}^{\rm min} \\ \hline
0.1 & 3.6\times10^{-6} \\
1 & 0.036 \\
2 & 0.57 \\
3 & 2.9
\end{array}
\end{equation}
Thus prompt Hagedorn reheating is easy when \(M_{s,S}\lesssim M_{s,A}\), but becomes difficult or impossible (which the last entry in \eqref{prompt_transfer_table} makes it evident) when the SM throat is much less strongly warped than the annihilation throat.
Hence, prompt transfer is easy if $M_{s, S} \lesssim M_{s, A}$, but difficult or impossible (as $B_{\text {vis }}\lesssim 1$) if the SM throat is less warped. Also, it is quite interesting to see that when the SM is more warped, the fraction of visible energy density dump can be quite low and still we can have a viable Hagedorn phase. So, from now on, we deal with the cases where the SM throat is more warped than the annihilation throat. 

We can also see how close the temperatures are, in these cases, to the Hagedorn temperature 
\begin{equation}\label{prompt closeness}
    \epsilon_H^{\text {prompt }}=\left[\frac{N^2 q_H^4 g_s}{4 \pi^3 m V_{\perp} B_{\text {vis }}}\left(\frac{M_{s,S}}{M_{s,A}}\right)^4\right]^{1 /2}.
\end{equation}
For \(q_H=1/\sqrt{2}\), \(m V_{\perp}=1\), \(g_s=0.05\), \(N=3\), and \(B_{\rm vis}=0.3\), Eq.~\eqref{prompt closeness} gives
\begin{equation}
\begin{array}{c|c|c}
M_{s,S}/M_{s,A}
&
\epsilon_H
&
T_{\rm rh}/T_H=1-\epsilon_H
\\ \hline
0.1 & 5.5\times10^{-4} & 0.99945 \\
0.5 & 1.4\times10^{-2} & 0.986 \\
1   & 5.5\times10^{-2} & 0.945
\end{array}
\end{equation}
Therefore, a more strongly warped SM throat reheats closer to the Hagedorn temperature. 

Using Eq.~\eqref{Duration}, the corresponding duration of the Hagedorn phase is
\begin{equation}
\begin{array}{c|c|c}
M_{s,S}/M_{s,A}
&
\epsilon_{H,i}
&
\Delta N_H
\\ \hline
0.1 & 5.5\times10^{-4} & 5.0 \\
0.5 & 1.4\times10^{-2} & 2.9 \\
1   & 5.5\times10^{-2} & 1.9
\end{array}
\end{equation}

So, the duration of the Hagedorn phase is short-lived. Substituting Eq.~\eqref{prompt closeness} into Eq.~\eqref{Delta Neff same throat}, one gets
\begin{equation}
\Delta N_{\mathrm{eff}}^{\mathrm{prompt}} \simeq 29.5 B_{\mathrm{dr}} B_{\mathrm{vis}}^{-4 / 3}\left(\frac{q_H^4 g_s}{4 \pi^3}\right)^{1 / 3}\left(\frac{M_{s, S}}{M_{s, A}}\right)^{4 / 3}
\end{equation}
Using $q_H=\frac{1}{\sqrt{2}}, \quad g_s=0.05$ and $B_{\text {vis }}=0.3$, 
\begin{equation}
\Delta N_{\mathrm{eff}}^{\mathrm{prompt}} \simeq 6.84 B_{\mathrm{dr}}\left(\frac{M_{s, S}}{M_{s, A}}\right)^{4 / 3} <0.3
\end{equation}
Thus,

\begin{tabular}{c|c|c}
$M_{s, S} / M_{s, A}$ & $\Delta N_{\text {eff }} / B_{\text {dr }}$ & $B_{\text {dr }}^{ \max }$ for $\Delta N_{\text {eff }}<0.3$ \\
\hline 0.1 & 0.317 & 0.95 \\
0.5 & 2.71 & 0.11 \\
1 & 6.84 & 0.044
\end{tabular}

\begin{figure}
    \centering
    \includegraphics[width=\linewidth]{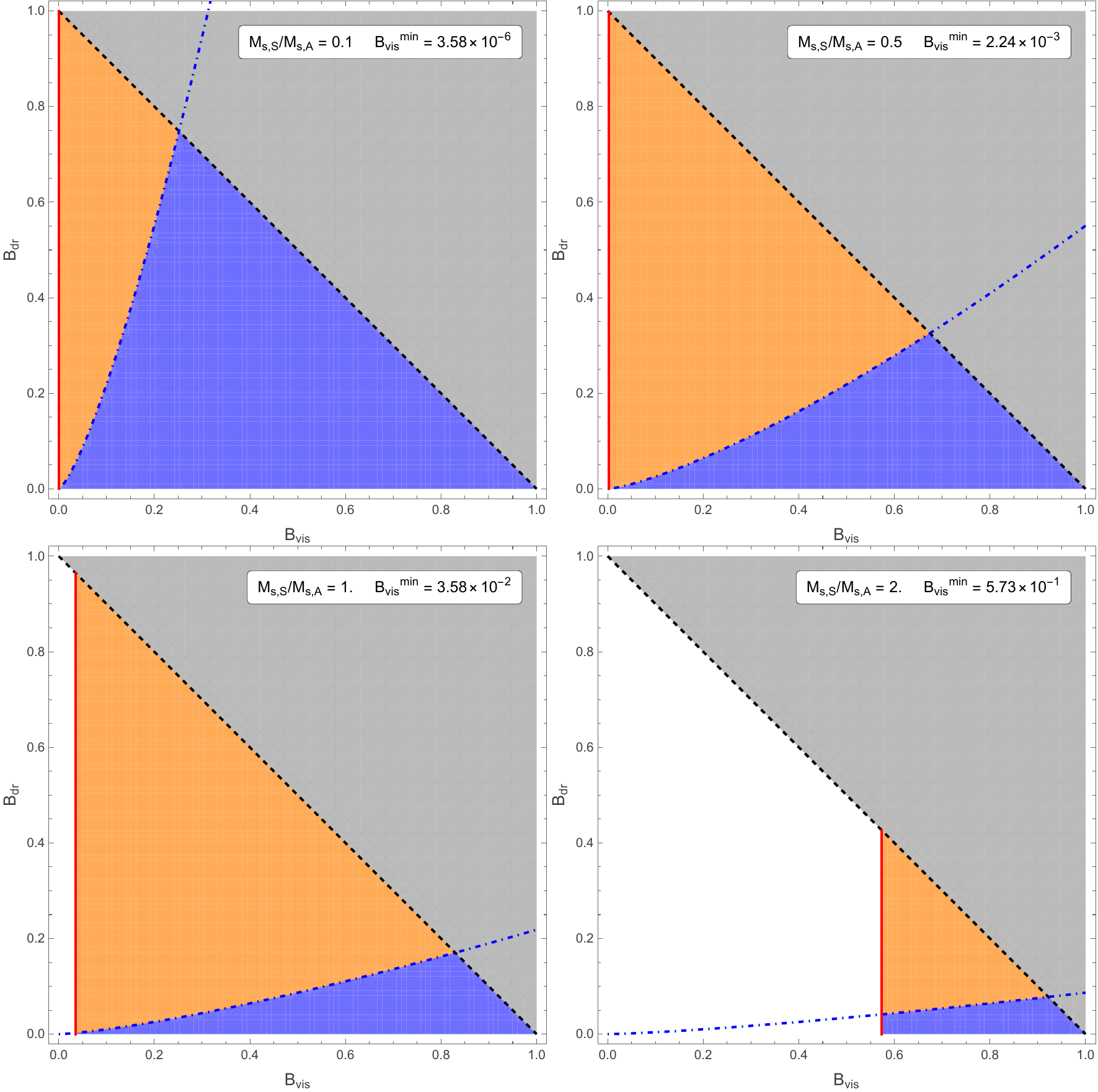}
    \caption{Prompt-transfer parameter space: Allowed regions in the $(B_{\rm vis},B_{\rm dr})$ plane for representative values of the ratio of local warped string scales $M_{s,S}/M_{s,A}$, with $g_s=0.05$, $N=3$, $\kappa_H=1$, and $q_H=1/\sqrt{2}$. The red vertical line denotes the minimum visible energy fraction required for Hagedorn entry, $B_{\rm vis}=B_{\rm vis}^{\rm min}$, so that the region to its right satisfies the Hagedorn condition. The blue dot-dashed curve denotes $\Delta N_{\rm eff}=0.3$, with the region below the curve satisfying the dark-radiation bound. The black dashed diagonal corresponds to $B_{\rm vis}+B_{\rm dr}=1$, while the gray region above it is unphysical. The blue-shaded region therefore satisfies Hagedorn entry, the bound $\Delta N_{\rm eff}<0.3$, and energy conservation simultaneously. The orange region below the diagonal is physical but fails either the Hagedorn-entry or dark-radiation conditions. Increasing $M_{s,S}/M_{s,A}$ progressively reduces the allowed region.}
    \label{fig:1}
\end{figure}

However, if $B_{\text {vis }}=0.3$ then $B_{\mathrm{dr}} \lesssim 0.7$. The resulting allowed parameter space in the prompt-transfer regime, together with the interplay between the Hagedorn-entry, dark-radiation, and energy-conservation constraints, is summarized in Figure~\ref{fig:1}. An important point to note is that not all fractions of the annihilation energy density will enter dark radiation or even visible sector due to inflationary throat being different and having certain tunneling probability into other sectors. 

\subsection{Regime 2: Delayed transfer}

We now consider the delayed-transfer regime, in which the energy stored in the annihilation throat \(A\) is not deposited into the SM throat immediately after brane-antibrane annihilation. This corresponds to
\begin{equation}
    \Gamma_{\rm tot}< H_{\rm end}.
\end{equation}
The energy reaches the SM throat when the Hubble scale has changed to some scale after inflation $H$:
\begin{equation}
    H\simeq \Gamma_{\rm tot}.
\end{equation}
Assuming that the \(A\)-throat energy dominates the total energy density until this time, the energy density available at deposition is
\begin{equation}
    \rho_{\rm dep}\simeq3M_p^2\Gamma_{\rm tot}^2.
\end{equation}
The visible open-string energy density is, therefore
\begin{equation}
    \rho_{\rm vis,open}^{(S)}=B_{\rm vis}\rho_{\rm dep}=3B_{\rm vis}M_p^2\Gamma_{\rm tot}^2,
\end{equation}
where
\begin{equation}
    B_{\rm vis}=\left(\frac{\Gamma_{A\to S}}{\Gamma_{\rm tot}}\right)B_{\rm vis}^{(S)}.
\end{equation}
Here, \(\Gamma_{\rm tot}\) controls the time at which the \(A\)-throat reservoir is depleted, while \(B_{\rm vis}\) controls the fraction of that available energy that ultimately appears as visible open strings in the throat \(S\).

Using the Hagedorn threshold condition in Eq.~\eqref{energy density condition}, the delayed-transfer Hagedorn condition becomes
\begin{equation}
    3B_{\rm vis}M_p^2\Gamma_{\rm tot}^2\gtrsim \kappa_HN^2M_{s,S}^4.
\end{equation}
Equivalently,
\begin{equation}
    \Gamma_{\rm tot}\gtrsim\Gamma_{\rm min},\qquad\Gamma_{\rm min}\equiv \sqrt{\frac{\kappa_HN^2}{3B_{\rm vis}}}\frac{M_{s,S}^2}{M_p}.
    \label{eq:gamma_miNelayed}
\end{equation}
Thus, if \(\Gamma_{\rm tot}<\Gamma_{\rm min}\), the universe has diluted too much before the SM throat is reheated, and the visible sector does not enter the Hagedorn regime.

The closeness to the Hagedorn temperature follows by equating \(\rho_{\rm vis,open}^{(S)}\) to the Hagedorn energy density in Eq.~\eqref{eq:hagedorn_energy_density}. This leads to
\begin{equation}
    \epsilon_H= \frac{Nq_H^2}{\sqrt{3\pi^2mV_{\perp}B_{\rm vis}}}\frac{M_{s,S}^2}{M_p\Gamma_{\rm tot}}.
    \label{eq:epsilon_delayed}
\end{equation}
Using Eq.~\eqref{eq:gamma_miNelayed}, this can be written more transparently as
\begin{equation}
    \epsilon_H=\frac{q_H^2}{\pi\sqrt{mV_{\perp}\kappa_H}}\frac{\Gamma_{\rm min}}{\Gamma_{\rm tot}}.
    \label{eq:epsilon_gamma_ratio}
\end{equation}
For $ q_H=\frac{1}{\sqrt2},\quad mV_{\perp}=1, \quad \kappa_H=1,$ and at the threshold \(\Gamma_{\rm tot}=\Gamma_{\rm min}\),
\begin{equation}
    T_{\rm rh}\simeq0.84\,T_{H,S},
\end{equation}
while larger \(\Gamma_{\rm tot}/\Gamma_{\rm min}\) corresponds to reheating closer to \(T_{H,S}\).

Using the duration estimate in Eq.~\eqref{Duration}, with \(\epsilon_{\rm exit}\sim1\), we find
\begin{equation}
    \Delta N_H\simeq\frac{2}{3} \ln\left(\frac{1}{\epsilon_H}\right) =\frac{2}{3} \ln\left[\frac{\pi\sqrt{mV_{\perp}\kappa_H}}{q_H^2}\frac{\Gamma_{\rm tot}}{\Gamma_{\rm min}} \right].
\end{equation}
For \(q_H=1/\sqrt2\), \(mV_{\perp}=1\), and \(\kappa_H=1\),
\begin{equation}
    \Delta N_H\simeq1.23+\frac{2}{3}\ln\left(\frac{\Gamma_{\rm tot}}{\Gamma_{\rm min}}\right).
\end{equation}
Representative values are:
\begin{equation}
\begin{array}{c|c|c}
\Gamma_{\rm tot}/\Gamma_{\rm min}
&
\epsilon_H
&
\Delta N_H
\\ \hline
1 & 0.159 & 1.23 \\
3 & 0.053 & 1.96 \\
10 & 0.0159 & 2.77 \\
30 & 5.3\times10^{-3} & 3.49 \\
100 & 1.59\times10^{-3} & 4.30
\end{array}
\end{equation}
Thus transfer only marginally above the Hagedorn threshold gives a short stringy phase, while faster transfer gives a longer near-Hagedorn epoch.

The dark-radiation contribution follows by substituting Eq.~\eqref{eq:epsilon_delayed} into Eq.~\eqref{Delta Neff same throat}. The factors of \(N\) and \(mV_{\perp}\) cancel and
for \(q_H=1/\sqrt2\), this reduces to
\begin{equation}
    \Delta N_{\rm eff}^{\rm delayed}\simeq6.0\,B_{\rm dr}B_{\rm vis}^{-4/3}\left(\frac{M_{s,S}^2}{M_p\Gamma_{\rm tot}}\right)^{2/3}.
\end{equation}
Taking \(N=3\), \(B_{\rm vis}=0.3\), \(\kappa_H=1\), \(mV_{\perp}=1\), and \(q_H=1/\sqrt2\), this becomes
\begin{equation}
    \Delta N_{\rm eff}^{\rm delayed}\simeq13.9\,B_{\rm dr}({\Gamma_{\rm tot}/\Gamma_{\rm min})}^{-2/3}.
\end{equation}
Therefore, we write
\begin{equation}
\begin{array}{c|c|c}
\Gamma_{\rm tot}/\Gamma_{\rm min}
&
\Delta N_{\rm eff}/B_{\rm dr}
&
B_{\rm dr}^{\rm max}
\quad
(\Delta N_{\rm eff}<0.3)
\\ \hline
1 & 13.9 & 0.022 \\
3 & 6.7 & 0.045 \\
10 & 3.0 & 0.10 \\
30 & 1.44 & 0.21 \\
100 & 0.64 & 0.47
\end{array}
\end{equation}
Thus, faster transfer suppresses \(\Delta N_{\rm eff}\), because the visible sector is reheated at a higher energy density and, therefore, closer to \(T_{H,S}\).

It remains to compare \(\Gamma_{\rm min}\) with \(H_{\rm end}\). The delayed Hagedorn regime exists only if the transfer rate can satisfy both
\begin{equation}
    \Gamma_{\rm min}\lesssim\Gamma_{\rm tot} <H_{\rm end}.
\end{equation}
Thus, \(\Gamma_{\rm min}\) must be below \(H_{\rm end}\); otherwise, the SM throat cannot enter the Hagedorn phase while the transfer is still delayed. Using $M_{s,A}\simeq3.4\times10^{13}\,{\rm GeV},  H_{\rm end}\simeq4.4\times10^9\,{\rm GeV}, M_p\simeq2.4\times10^{18}\,{\rm GeV},$ and for \(N=3\), \(\kappa_H=1\), and \(B_{\rm vis}=0.3\), Eq.~\eqref{eq:gamma_miNelayed} gives
\begin{equation}
    \Gamma_{\rm min}\simeq1.5\times10^9\left(\frac{M_{s,S}}{M_{s,A}}\right)^2{\rm GeV}.
\end{equation}
Therefore, the delayed Hagedorn window requires
\begin{equation}
    \frac{M_{s,S}}{M_{s,A}}\lesssim 1.7.
\end{equation}
Thus, the SM throat cannot be much less warped than the annihilation throat. A more strongly warped SM throat, \(M_{s,S}\ll M_{s,A}\), makes the condition much easier because \(\Gamma_{\rm min}\propto M_{s,S}^2\).

\begin{figure}
    \centering
    \includegraphics[width=0.8\linewidth]{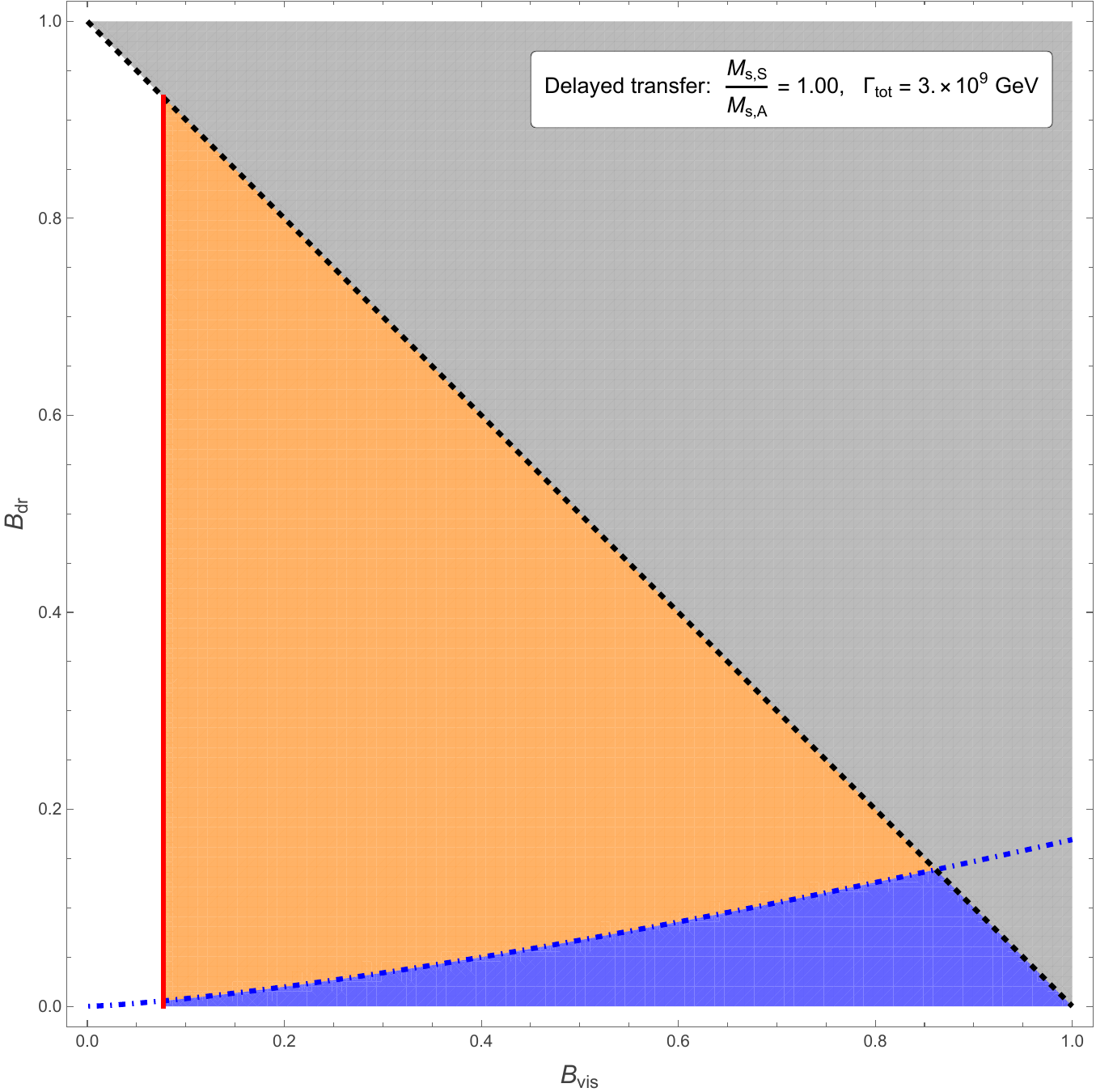}
    \caption{Parameter space for delayed transfer: Allowed regions in the $(B_{\rm vis},B_{\rm dr})$ plane for the benchmark $M_{s,S}/M_{s,A}=1$ and $\Gamma_{\rm tot}=3\times10^{9}\,{\rm GeV}$, with $\Gamma_{\rm tot}<H_{\rm end}$ and the remaining parameters chosen as in the prompt-transfer analysis. The red dashed line denotes the Hagedorn-entry boundary $\Gamma_{\rm tot}=\Gamma_{\rm min}$; the white region to its left satisfies energy conservation but fails to enter the Hagedorn regime because $\Gamma_{\rm tot}<\Gamma_{\rm min}$. The blue dot-dashed curve denotes $\Delta N_{\rm eff}=0.3$. The orange region enters the Hagedorn regime but violates the dark-radiation bound, whereas the blue region satisfies Hagedorn entry, $\Delta N_{\rm eff}<0.3$, and energy conservation simultaneously. The black solid diagonal denotes $B_{\rm vis}+B_{\rm dr}=1$, and the gray region above it is unphysical.}
    \label{fig:placeholder}
\end{figure}

Numerically,
\begin{equation}
\begin{array}{c|c|c}
M_{s,S}/M_{s,A}
&
\Gamma_{\rm min}\ [{\rm GeV}]
&
\text{delayed Hagedorn window}
\\ \hline
0.1 & 1.5\times10^7 & \text{broad} \\
0.3 & 1.4\times10^8 & \text{broad} \\
0.5 & 3.8\times10^8 & \text{moderate} \\
1 & 1.5\times10^9 & \text{narrow} \\
1.5 & 3.4\times10^9 & \text{very narrow} \\
2 & 6.1\times10^9 & \text{none}
\end{array}
\end{equation}
Hence, delayed Hagedorn reheating is naturally preferred when the SM throat is more strongly warped than the annihilation throat. For comparable throats, the window exists but is narrow, while for \(M_{s,S}\gtrsim2M_{s,A}\) the required Hagedorn-entry rate already exceeds \(H_{\rm end}\), so the transfer cannot be both delayed and Hagedorn-producing. The corresponding allowed parameter space in the delayed-transfer regime, and the interplay between the Hagedorn-entry, dark-radiation, and energy-conservation constraints, is summarized in Figure~\ref{fig:placeholder}.
\section{Conclusion and Remarks}
In this work, we studied the possibility that the endpoint of perturbatively stabilised D3/\(\overline{\mathrm{D3}}\) inflation is not described immediately by an ordinary radiation bath, but instead passes through an intermediate open-string Hagedorn phase. The perturbatively stabilised construction of Ref.~\cite{Cicoli:2024bwq} provides a controlled setting in which brane-antibrane inflation can be studied. In this setting, the annihilation energy is of the order of the warped brane tension, \(\rho_{\rm ann}^{(A)}\simeq (4\pi/g_s)M_{s,A}^4\), and the density is therefore naturally above the local string scale in the annihilation throat. Whether the visible sector enters a Hagedorn regime depends on the visible energy fraction and on the local string scale of the Standard Model throat. In the case where inflation and SM is realized in the same-throat, we found that the Hagedorn condition is mild: for weak string coupling, a few to ten percent of the annihilation energy deposited into surviving visible open strings is enough to reach the open-string Hagedorn phase. This Hagedorn phase in turn, like \cite{Frey:2021jyo}, suppresses $\Delta N_{\rm eff}$. In the different-throat case, the result depends crucially on inter-throat transfer. Prompt transfer is efficient when the SM throat is at least as strongly warped as the annihilation throat \(M_{s,S}\lesssim M_{s,A}\), while delayed transfer requires the transfer rate to exceed the minimum Hagedorn-entry rate before cosmological dilution becomes too large. The preferred regime for visible Hagedorn reheating for delayed transfer is one in which the SM local warped string scale is lower than, or at most comparable to, the annihilation-throat string scale.

\section*{Acknowledgements}
ARK is supported by the Czech Science Foundation GA\v{C}R grant "One loop in ten and eleven dimensions"(GA26-22343S) and also acknowledges support from BRAC University BRACURSGI25014. ARK thanks KU Leuven physics department and especially Thomas Van Riet for hosting at Leuven where part of this work was done. DC would like to thank the warm hospitality of Indian Statistical Institute Kolkata and IISER-Pune where part of this research work has been conducted.

The generative AI \texttt{ChatGPT-Plus} is used to style the file as well as for grammatical improvements.

\bibliographystyle{apsrev4-2}
\bibliography{biblio}

\end{document}